\documentclass[twocolumn,showpacs,superscriptaddress,pra]{revtex4}
\usepackage{graphicx}
\usepackage{amsmath}
\usepackage{amssymb}
\usepackage{latexsym}
\usepackage{array}
\usepackage{amsfonts}
\usepackage{mathrsfs}
\usepackage{verbatim}
\usepackage{multirow}

\begin{document}

\title{Quantum-mechanical machinery for rational decision-making in classical guessing game}

\author{Jeongho Bang}\email{jbang@gist.ac.kr}
\affiliation{Center for Photon Information Processing, and School of Information and Communications, Gwangju Institute of Science and Technology, Gwangju, Republic of Korea}
\affiliation{Department of Physics, Hanyang University, Seoul 133-791, Korea}
\affiliation{Institute of Theoretical Physics and Astrophysics, University of Gda\'{n}sk, 80-952 Gda\'{n}sk, Poland}

\author{Junghee Ryu}
\affiliation{Institute of Theoretical Physics and Astrophysics, University of Gda\'{n}sk, 80-952 Gda\'{n}sk, Poland}

\author{Marcin Paw\l{}owski}
\affiliation{Institute of Theoretical Physics and Astrophysics, University of Gda\'{n}sk, 80-952 Gda\'{n}sk, Poland}

\author{B. S. Ham}\email{bham@gist.ac.kr}
\affiliation{Center for Photon Information Processing, and School of Information and Communications, Gwangju Institute of Science and Technology, Gwangju, Republic of Korea}

\author{Jinhyoung Lee}\email{hyoung@hanyang.ac.kr}
\affiliation{Department of Physics, Hanyang University, Seoul 133-791, Korea}


\received{\today}

\begin{abstract}
In quantum game theory, one of the most intriguing and important questions is, ``Is it possible to get {\em quantum} advantages without any modification of the {\em classical} game?'' The answer to this question so far has largely been negative. So far, it has usually been thought that a change of the classical game setting appears to be unavoidable for getting the quantum advantages. However, we give an affirmative answer here, focusing on the decision-making process (we call `reasoning') to generate the best strategy, which may occur internally, e.g., in the player's brain. To show this, we consider a classical guessing game. We then define a one-player reasoning problem in the context of the decision-making theory, where the machinery processes are designed to simulate classical and quantum reasoning. In such settings, we present a scenario where a rational player is able to make better use of his/her weak preferences due to quantum reasoning, without any altering or resetting of the classically defined game. We also argue in further analysis that the quantum reasoning may make the player fail, and even make the situation worse, due to any inappropriate preferences.
\end{abstract}


\maketitle

\newcommand{\bra}[1]{\left<#1\right|}
\newcommand{\ket}[1]{\left|#1\right>}
\newcommand{\abs}[1]{\left|#1\right|}
\newcommand{\expt}[1]{\left<#1\right>}
\newcommand{\braket}[2]{\left<{#1}|{#2}\right>}
\newcommand{\ketbra}[2]{\left|{#1}\right>\left<{#2}\right|}
\newcommand{\commt}[2]{\left[{#1},{#2}\right]}
\newcommand{\tr}[1]{\mbox{Tr}{#1}}


\section{Introduction}\label{sec:1}

Game theory, which is a very well established discipline in mathematics \cite{Julio10}, has successfully been applied to various fields, such as social science \cite{Camerer03,Castellano09}, evolutions in biology \cite{Weibull97}, and economics \cite{Aumann92}. At the abstract level, game theory mainly deals with legitimate strategies and scores of the players. Thus, a game is defined by the strategies on one hand, and by a specification of how to evaluate the players' game scores on the other hand. Recently, physicists have been attempting to generalize the game into a new scenario finding common theoretical properties between the game and quantum theory \cite{Meyer99,Eisert99,Eisert00,Piotrowski03,Guo08}. Of particular interest to this generalization is to study whether it is possible to replace the classical strategy with a quantum strategy for getting quantum advantages, if any \cite{Lee03}. The quantum advantages from such a generalization have been found to be relevant to these games. For example, consider the ``penny-flip game'' \cite{Meyer99}, where two players take turns choosing whether or not to flip a penny inside a box, and the starting player opens the box to identify if the penny is flipped from its starting position or not. Here, if one player can adopt a quantum penny, then he/she has a better chance of winning assisted by quantum superposition. Another celebrated example is the ``Prisoner's game'' \cite{Eisert99,Eisert00}, where two players face a dilemma, since acting rationally for their own interests would result in a collectively worse outcome. But this dilemma can also be solved by adopting quantum strategies that the players can realize. Most recently, some new game scenarios have been conceived, that establish a strong link to communication complexity \cite{Muhammad14} and Bell-inequality engaging the nonlocality \cite{Brunner13,Pappa15}.

Following up on the successes of the previous studies of the quantum game, we also plan to explore a positive role of the `quantum' in a classically designed game. In particular, we consider the following question, ``Is it possible to get quantum advantages without any quantum modification?'' This question is important because nearly all games are allowed to have the advantages due to ``quantum strategies'', but it has usually been thought that one inevitably needs to change the original form of the classical game to enjoy the quantum advantages \cite{SJvanEnk02,Aharon08}. Therefore, the answer to the aforementioned question has been negative. However, here we find an affirmative answer, focusing on --- substantially different from the earlier approaches  --- the decision-making (we call ``reasoning'' hereafter) of the rational player. To show this, we design a classical two-player game, called the Secret-Bit Guessing Game, where one player named Bob attempts to guess the secret bits of the other player, Alice. For this game, we map out two parallel ways of Bob's reasoning to choose his best answering strategies: one is classical probabilistic, and the other is quantum. Each reasoning that is drawn in Bob's brain is modeled as a machinery process for systematic analysis and fair comparison. On the basis of the payoff-function analysis, we explicitly show that the quantum reasoning can be more advantageous without changing the classical setting of the game. This is because the rational player, Bob, can make better use of his weak preferences, faithfully dealing with quantum superposition. However, we also argue in further analysis that the quantum reasoning may frustrate Bob, and even make the situation worse due to malicious hints that can lead Bob to have the wrong preferences.

\section{Preliminary}\label{sec:2}

\subsection{A secert-bit guessing game}

Firstly, we consider a secret-bit guessing game \cite{Chung01,Lungo05}. In the game, Alice generates two bits and keeps them on her memory $M_x$ ($x=0,1$). Here we note that the identity of the bits are in $M_x$, regardless of whether anyone can access it or not (i.e., the secret bits are classical). Then, the other player Bob chooses his answering strategies $u_\text{Bob} \in \{0,1\}$ to guess the secret bits enveloped by Alice, considering four possible strategies $u_\text{Alice}(x)$ ($x=0,1$) that Alice may have. Specifically, 
\begin{eqnarray}
M_x \leftarrow
\left\{
\begin{array}{ll}
\left[\boldsymbol\tau.{\bf 1}\right] & u_\text{Alice}(x)=0, \\
\left[\boldsymbol\tau.{\bf 2}\right] & u_\text{Alice}(x)=x,  \\
\left[\boldsymbol\tau.{\bf 3}\right] & u_\text{Alice}(x)=1 \oplus x, \\
\left[\boldsymbol\tau.{\bf 4}\right] & u_\text{Alice}(x)=1, 
\end{array}
\right.
\label{eq:Alice_sbit}
\end{eqnarray}
where `$\oplus$' denotes modulo-$2$ addition. Here, if Bob's answer is correct (i.e., $u_\text{Bob} = u_\text{Alice}$) for a given $x$, Bob wins a single-point $\frac{1}{2}$ and Alice loses the same single-point. Otherwise, in case Bob gives wrong answer (i.e., $u_\text{Bob} \neq u_\text{Alice}$), Alice and Bob get the single-points $\frac{1}{2}$ and $-\frac{1}{2}$, respectively. This game is thus defined as ${\cal G}={\cal G}(S, P)$, where $S$ and $P$ denote the (non-empty) sets of the players' strategies ($u_\text{Alice}$, $u_\text{Bob}$) and game scores ($\xi_\text{Alice}$, $\xi_\text{Bob}$), respectively. Noting that Bob makes two answers for $M_x$ ($x=0,1$), the possible game scores for Alice and Bob after one game are made by adding the two single-points, and thus $\xi_\text{Alice},\xi_\text{Bob} \in \{-1, 0, 1\}$. Note further that $\xi_{\text{Alice}}$ is equal to $-\xi_{\text{Bob}}$, or equivalently, $\xi_{\text{Alice}} + \xi_{\text{Bob}}=0$; i.e., our game is zero-sum \cite{Julio10}. 


\subsection{One-player reasoning problem}

Whereas in previous studies the strategies have usually been generalized in a quantum regime, our primary concern here is with the reasoning process. In particular, we would like to investigate if a quantum reasoning can yield a higher winning average compared to the classical ones even in a fully classical game. We now turn our attention to Bob's reasoning to make his valid answering strategies $u_\text{Bob}$ for $x=0,1$. We define this a ``One-Player Reasoning Problem.'' 

To deal with this problem, we design a process of Bob's reasoning by introducing a one-bit Boolean function,
\begin{eqnarray}
u_\text{Bob}(x)=r_0 \oplus r_1 x, 
\label{eq:h_f}
\end{eqnarray}
where $r_0, r_1 \in \{0, 1\}$. Then, Bob's reasoning is nothing but the process of making the output $u_\text{Bob}(x)$ for a given $x \in \{0,1\}$, depending on the coefficients $(r_0, r_1) \in R$. Note that the function in Eq.~(\ref{eq:h_f}) can generate all possible sets [$\boldsymbol\tau.{\bf 1}$]-[$\boldsymbol\tau.{\bf 4}$] of $u_\text{Alice}(x)$ in Eq.~(\ref{eq:Alice_sbit}). Here we consider the concept of a {\em hint} given from, e.g., a helper \footnote{Note that the hints are referred to as the classical information, as the real-world players recognize the measured information.}, which allows Bob to have (`weak' or possibly `strong') {\em preferences} over $R$. Thus, we can formulate our problem ($R$, $\succsim$) with Bob's preferences and alternatives $R$ in the context of the theory of decision-making \cite{Julio10}. We note that the hints are presented in abstract form. We assume an ``interpretation function'' that quantifies his own preferences, such that \footnote{One may think that such a function appears to be anomaly or artificial. However, in a conventional computation, we frequently encounter such a situation that an abstract knowledge should be applied to the designed operations as a certain value. Actually, most of the non-trivial operation, e.g., giving an answer ``yes/no'' as a signal ``0/1'', or marking a specific property as a specific number, etc., involves such a interpretation function implicitly.}
\begin{eqnarray}
\{ \text{Pr}(r_0 \rightarrow k), \text{Pr}(r_1 \rightarrow k') \} \in H ~ (k, k' = 0,1),
\label{eq:quantified_hint}
\end{eqnarray}
where $\text{Pr}(r_j \rightarrow k)$ denotes the probability of choosing ``$r_j \rightarrow k$'' ($j,k=0,1$), and $H$ denotes the possible set of those probabilities. Here, $\text{Pr}(r_j \to k \oplus 1) = 1 - \text{Pr}(r_j \to k)$. Thus if $\text{Pr}(r_j \rightarrow k) \ge \frac{1}{2}$, Bob wants to choose ``$r_j \to k$'', at least as much as ``$r_j \to k \oplus 1$'' (i.e., ``$r_j \to k$'' $\succsim$ ``$r_j \to k \oplus 1$''), and vice versa. More specifically, we write $\text{Pr}(r_j \rightarrow k)$ as
\begin{eqnarray}
\text{Pr}(r_j \rightarrow k) &=& \frac{1}{2} + (-1)^k \alpha_j, 
\label{eq:hint_alpha}
\end{eqnarray}
where $\alpha_j \in [-\frac{1}{2}, \frac{1}{2}]$ is defined as a factor to represent the bias of Bob's preferences. 

\begin{figure}[t]
\centering
\includegraphics[width=0.35\textwidth]{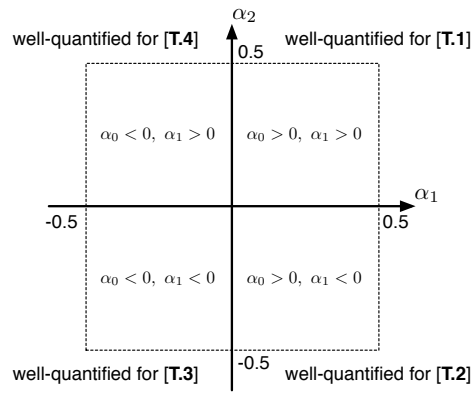}
\caption{(Color online) For each set of Alice's strategies $u_\text{Alice}(x)$ ($x=0,1$), we specify the regions of well-quantified probabilities of Bob's preferences in the space of ($\alpha_0$, $\alpha_1$).}
\label{fig:good_advice}
\end{figure}

\begin{figure}[t]
\centering
\includegraphics[width=0.45\textwidth]{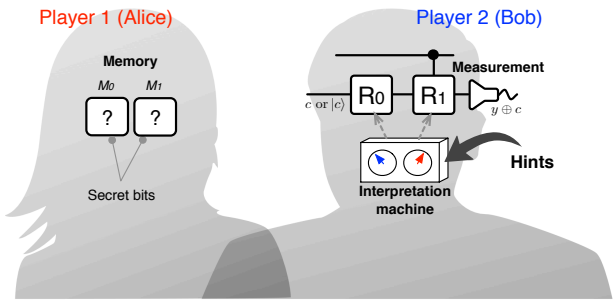}
\caption{(Color online) Schematic picture of our game setting. Alice sets the two secret bits into her memory $M_x$ ($x=0,1$) and Bob attempts to guess them. In this game, we define one-player (Bob's) reasoning problem with a certain set of probabilities of the player's own preferences, as in Eq.~(\ref{eq:quantified_hint}). Here, we replace the reasoning to the process of a machinery that consists of the corresponding internal devices involved (e.g., one inside the player's brain). See main text in Sec.~\ref{sec:3}.}
\label{fig:game}
\end{figure}

Here, if the direction of $\boldsymbol\alpha = (\alpha_0, \alpha_1)^T$ is appropriately assigned, we say that the probabilities in Eq.~(\ref{eq:quantified_hint}) are {\em well-quantified}, where $\boldsymbol\alpha$ was defined as a vector on the two-dimensional space of ($\alpha_1$, $\alpha_2$). For example, if Alice's strategies $u_\text{Alice}(x)$ are [$\boldsymbol\tau.{\bf 1}$], the well-quantified probabilities are characterized with the directional condition ($\alpha_1 > 0$ and $\alpha_2 > 0$) (see Fig.~\ref{fig:good_advice} for all the cases). Bob is supposed to perform his reasoning, believing his ability to interpret the given hints. A schematic picture of our game is presented in Fig.~\ref{fig:game}.

\section{Two parallel reasoning processes: Classical probabilistic, and quantum}\label{sec:3}

In order to perform a more systematic analysis, we replace the reasoning to a machinery process, which may occur, e.g., in Bob's brain. To this end, we consider a fledged computing module, as depicted in Fig.~\ref{fig:game}, to simulate the reasoning process of Eq.~(\ref{eq:h_f}) \cite{Younes04}. This computing module consists of two one-way channels ${\cal C}_x$ and ${\cal C}_y$, where ${\cal C}_x$ transmits the classical signals of the memory number $x$, and ${\cal C}_y$ deals with the signals of Bob's strategies $u_\text{Bob}(x)$ ($x=0,1$). Two probabilistic logic gates $R_0$ and $R_1$ are also placed in ${\cal C}_y$, but note that $R_1$ acts conditioned on the input $x$ in ${\cal C}_x$ being $1$. Bob's strategies $u_\text{Bob}(x)$ are identified by the measurement at the end of ${\cal C}_y$. We here introduce another internal machine, to be called an interpretation machine, to generate the probabilities $\text{Pr}(r_j \to k)$ ($j,k=0,1$) of Bob's preferences.

With this general model of reasoning machinery, we assume that Bob can make two different types of reasoning: {\em Classical probabilistic} and {\em quantum}. Here, we note that the signals $x$ in ${\cal C}_x$ should be classical even in the case of quantum reasoning, as it is regarded as an element of the game. Thus, we do not need to consider any additional internal process to convert the classical information to the quantum information in the reasoning, or vice versa \cite{Aharon08}. This assumption is not trivial, as we have to make a fair comparison of the two reasonings, independently from the classically designed game \footnote{Here, if Bob is allowed to convert the classical input $x$ to the quantum, e.g., $\ket{x}$ in his quantum reasoning and thus to use the quantum properties of them, it is unfair and tricky because it is, in fact, the use of the quantum game element.}.

\begin{figure}[t]
\centering
\includegraphics[width=0.42\textwidth]{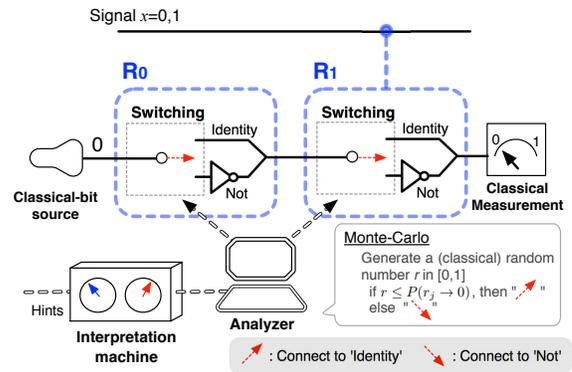}
\caption{(Color online) A realizable and concrete setting of Bob's classical probabilistic reasoning. In such a setting, the analyzer (or the controller) receives the quantified probabilities from the interpretation machine, and performs the Monte-Carlo method by generating a (classical) random number $r \in [0,1)$. Here, if the randomly generated $r$ is smaller (or larger) than $\text{Pr}(r_j \rightarrow 0)$, the switching device of $R_j$ connects the incoming signal to `Identity' (or `{\small NOT}'). Bob {\em identifies} a value of $u_\text{Bob}(x)$ for given $x$ in the classical measurement.}
\label{fig:classical_a}
\end{figure}

$1$) {\em Classical probabilistic reasoning.} -- Firstly, let us assume that the signals of ${\cal C}_y$ are classical and the logic gates $R_j$ ($j=0,1$) act according to the probabilistic rule, to be either ``$\openone$'' with the probability $\text{Pr}(r_j \rightarrow 0)$ or ``{\small NOT}'' with the probability $\text{Pr}(r_j \rightarrow 1)$. In this case, $R_0$ and $R_1$ are {\em classical probabilistic gates}. Such a probabilistic instruction results in better computational outputs in a heuristic manner \cite{Shachter92}, and could allow a reasonable comparison with the unitary gates adopted in the quantum reasoning (as described later). To make the strategy of the answer $u_\text{Bob}(x)$ for a given $x$, the final value $0$ or $1$ passing through the gates $R_j$ ($j=0,1$) is identified by the classical measurement. In Fig.~\ref{fig:classical_a}, we sketch a realizable and concrete implementation of such a machinery process for the classical probabilistic reasoning.

\begin{figure}[t]
\centering
\includegraphics[width=0.42\textwidth]{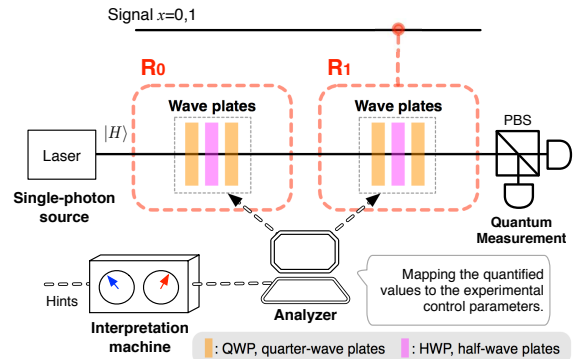}
\caption{(Color online) A setting of Bob's quantum reasoning. Here we consider a linear optical implementation, where the signals of ${\cal C}_y$ are encoded as polarized single-photon states $\ket{H}$ and $\ket{V}$ ($H=0$ and $V=1$). The unitary gates $\hat{R}_j$ ($j=0,1$) are realized by a set of wave plates (QWP-HWP-QWP) for the polarized photon \cite{Damask04}. The analyzer maps the quantified probabilities in Eq.~(\ref{eq:quantified_hint}) to the control parameters of the wave-plates, using Eq.~(\ref{eq:q_U}) and Eq.~(\ref{eq:Delta}). Then, Bob's brain performs the quantum measurement on the final output photon.}
\label{fig:quantum_a}
\end{figure}

$2$) {\em Quantum reasoning.} -- On the other hand, Bob can also follow the quantum reasoning, where the crucial part of the computing module, including the channel ${\cal C}_y$, logic gates $R_j$ ($j=0,1$), and measurement device, are quantum. In such a case, each of the gates $R_j$ ($j=0,1$) is to be a unitary transformation, defined as
\begin{eqnarray}
\hat{R}_j = \begin{pmatrix} \sqrt{\text{Pr}(r_j \to 0)} & e^{i \phi_j} \sqrt{\text{Pr}(r_j \to 1)} \\ e^{-i \phi_j} \sqrt{\text{Pr}(r_j \to 1)} & -\sqrt{\text{Pr}(r_j \rightarrow 0)} \end{pmatrix},
\label{eq:q_U}
\end{eqnarray}
which also leaves and flips the states $\ket{0}$ and $\ket{1}$ with the probability $\text{Pr}(r_j \to 0)$ and $\text{Pr}(r_j \to 1)$, respectively. However, it should be noted that the unitary gate $\hat{R}_j$ has an additional degree of freedom, i.e., {\em quantum phase} $\phi_j$, to exhibit the genuine property of the quantum superposition \footnote{Actually, it is well-known that the quantum phases and their control are inevitably required to fully enjoy the computational power in a conventional scheme of quantum computation \cite{Barenco95}. Here we also would like to note that we can leave room for the possibility that the entanglement could be engaged in the quantum advantages \cite{Meyer00,Du00,Kenigsberg06,Li11}. Actually, the computing module designed in the quantum action cannot generate the entanglement because the signals of ${\cal C}_x$ are deterministic or classical.}. It allows the {\em rational} player, Bob, to explore an additional rule for setting the phases $\phi_j$ ($j=0,1$) to maximize his winning averages. In our game, Bob additionally uses the directional condition of $\boldsymbol{\alpha}$, which could not considered in the classical probabilistic reasoning. More specifically, Bob's brain sets the phases $\phi_j$ ($j=0,1$), according to
\begin{eqnarray}
\left\{
\begin{array}{lll}
(i) & \Delta = 0 & \text{if}~\alpha_0 \alpha_1 > 0, \\
(ii) & \Delta = \frac{1}{2} & \text{if}~\alpha_0 \alpha_1=0, \\
(iii) & \Delta = 1 & \text{if}~\alpha_0 \alpha_1 < 0,
\end{array}
\right.
\label{eq:Delta}
\end{eqnarray}
where $\Delta = \phi_1 - \phi_0$. These rules were made to maximize Bob's winning averages. Here, the case ($i$) describes the situation that Bob's brain, internally, chooses $\Delta=0$ when $\boldsymbol\alpha$ contains the directional conditions ($\alpha_0 > 0$, $\alpha_1 > 0$) or ($\alpha_0 < 0$, $\alpha_1 < 0$), which are toward [$\boldsymbol\tau.{\bf 1}$] and [$\boldsymbol\tau.{\bf 3}$], respectively. In the case of ($iii$), $\Delta$ is set to be $1$ with $(\alpha_0 > 0, \alpha_1 < 0)$ or $(\alpha_0 < 0, \alpha_1 > 0)$ whose directions are toward [$\boldsymbol\tau.{\bf 2}$] and [$\boldsymbol\tau.{\bf 4}$], respectively. However, in the case of ($ii$), i.e., when $\alpha_0=0$ or $\alpha_1=0$, it is not possible to find any useful setting, as a feasible direction of $\boldsymbol\alpha$ cannot be sured. At the final step, Bob's brain performs the quantum measurement on the final state to get $u_\text{Bob}(x)$. Note that, in the view of the intrinsic probabilistic nature of the quantum system, the final state does not result in a definite or predictable outcome value. In Fig.~\ref{fig:quantum_a}, a schematic example of such a quantum reasoning procedure is sketched in the linear-optical regime.

Here, we briefly comment that the two settings described above are parallel in the sense that the operation of the logic gates are comparable in each reasoning. Note that, for the number of games, the operations of the classical probabilistic gates $R_j$ ($j=0,1$) can also be represented by a stochastic evolution matrix as
\begin{eqnarray}
R_j = \begin{pmatrix} \text{Pr}(r_j \to 0) & \text{Pr}(r_j \to 1) \\ \text{Pr}(r_j \to 1) & \text{Pr}(r_j \to 0)  \end{pmatrix},
\label{eq:c_S}
\end{eqnarray}
where the matrix elements may provide the candidate transition probabilities which describe the dynamics of unitary transformation of Eq.~(\ref{eq:q_U}) \footnote{Of course, this is not the case. See Ref. \cite{Schack99}.}. 

\section{Analysis of Bob's average scores achievable from the two reasonings}\label{sec:4}

A crucial task in game theory is to investigate a function $f_{\$}$, which determines the average scores of the players over the number of games. In our game, such a function $f_{\$}$, named the payoff function, can be defined by
\begin{eqnarray}
f_{\$} : S \times H \times A \rightarrow \left(\Xi_\text{Alice} \in {\mathbb R}, \Xi_\text{Bob} \in {\mathbb R}\right),
\end{eqnarray}
where $\Xi_\text{Alice}$ and $\Xi_\text{Bob}$ denote the total average scores of Alice and Bob, respectively, and $A$ denotes the set of possible reasonings. As mentioned before, our game is a two-player zero-sum game, so it is sufficient to analyze the score of one of the players. We thus focus on the average score $\Xi_\text{Bob}$ of Bob throughout the work. 

More specifically, the total average score $\Xi_{\text{Bob}}$ can be evaluated as
\begin{eqnarray}
\Xi_{\text{Bob}} = \frac{1}{4}\sum_{\tau=1}^{4} \overline{\xi}_{\text{Bob},\tau},
\label{eq:payoff_gen}
\end{eqnarray}
where it is assumed that Alice chooses her bits at random. Here, $\overline{\xi}_{\text{Bob},\tau}$ ($\tau=1,2,3,4$) is also defined as the score averaged for a specific set of $u_\text{Alice}(x)$ ($j=0,1$), 
\begin{widetext}
\begin{eqnarray}
\overline{\xi}_{\text{Bob},\tau} = \sum_{x=0,1} \left( \frac{1}{2}\text{Pr}\left(u_\text{Bob}(x)=u_\text{Alice}(x)\right) - \frac{1}{2}\text{Pr}\left(u_\text{Bob}(x) \neq u_\text{Alice}(x)\right) \right),
\label{eq:payoff_av}
\end{eqnarray}
where $\text{Pr}({u_\text{Bob}(x)=u_\text{Alice}(x)})$ and $\text{Pr}({u_\text{Bob}(x) \neq u_\text{Alice}(x)})$ are the probabilities that Bob's answer is correct and incorrect for a given $x$, respectively, and $\tau$ denotes the index of the possible sets [$\boldsymbol\tau.{\bf 1}$]-[$\boldsymbol\tau.{\bf 4}$] of Alice's strategies $u_\text{Alice}(x)$. For our later analysis, we here rewrite $\overline{\xi}_{\text{Bob},\tau}$, for each $\tau$, as
\begin{eqnarray}
\overline{\xi}_{\text{Bob},1} &=& \frac{1}{2}\left( \text{Pr}(u_\text{Bob}=0|x=0) + \text{Pr}(u_\text{Bob}=0|x=1) - \text{Pr}(u_\text{Bob}=1|x=0) - \text{Pr}(u_\text{Bob}=1|x=1) \right), \nonumber \\
\overline{\xi}_{\text{Bob},2} &=& \frac{1}{2}\left( \text{Pr}(u_\text{Bob}=0|x=0) + \text{Pr}(u_\text{Bob}=1|x=1) - \text{Pr}(u_\text{Bob}=1|x=0) - \text{Pr}(u_\text{Bob}=0|x=1) \right), \nonumber \\
\overline{\xi}_{\text{Bob},3} &=& \frac{1}{2}\left( \text{Pr}(u_\text{Bob}=1|x=0) + \text{Pr}(u_\text{Bob}=0|x=1) - \text{Pr}(u_\text{Bob}=0|x=0) - \text{Pr}(u_\text{Bob}=1|x=1) \right), \nonumber \\
\overline{\xi}_{\text{Bob},4} &=& \frac{1}{2}\left( \text{Pr}(u_\text{Bob}=1|x=0) + \text{Pr}(u_\text{Bob}=1|x=1) - \text{Pr}(u_\text{Bob}=0|x=0) - \text{Pr}(u_\text{Bob}=0|x=1) \right),
\label{eq:payoff_av_tau}
\end{eqnarray}
\end{widetext}
where $\text{Pr}(u_\text{Bob}|x)$ ($x=0,1$) is the probability that Bob identifies $u_\text{Bob}(x)$, given the memory number $x$. In the following, we shall analyze the total average score $\Xi_{\text{Bob}}$ achievable from each of the two reasonings. 

\begin{figure}[t]
\centering
\includegraphics[width=0.26\textwidth]{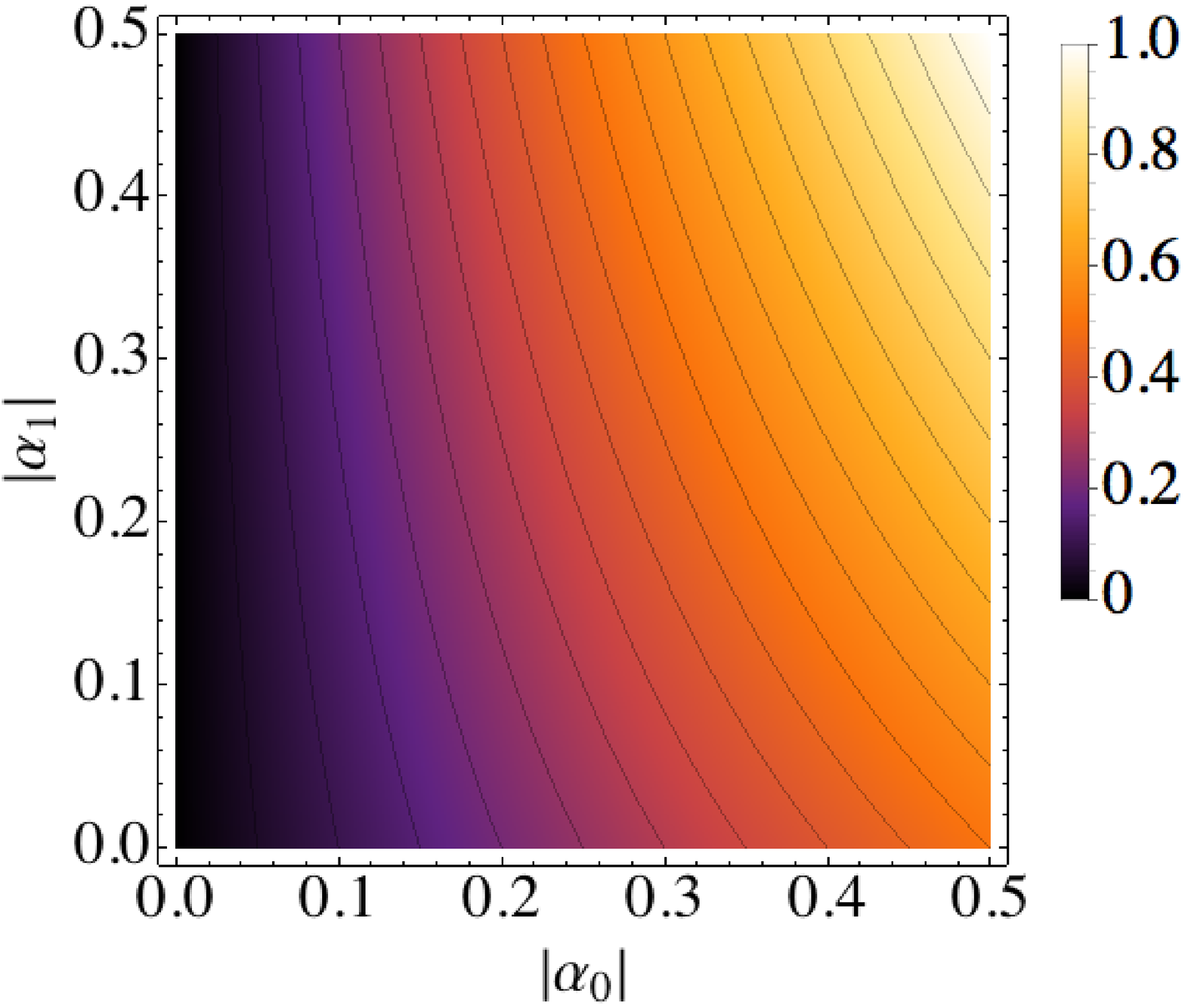}
\includegraphics[width=0.2\textwidth]{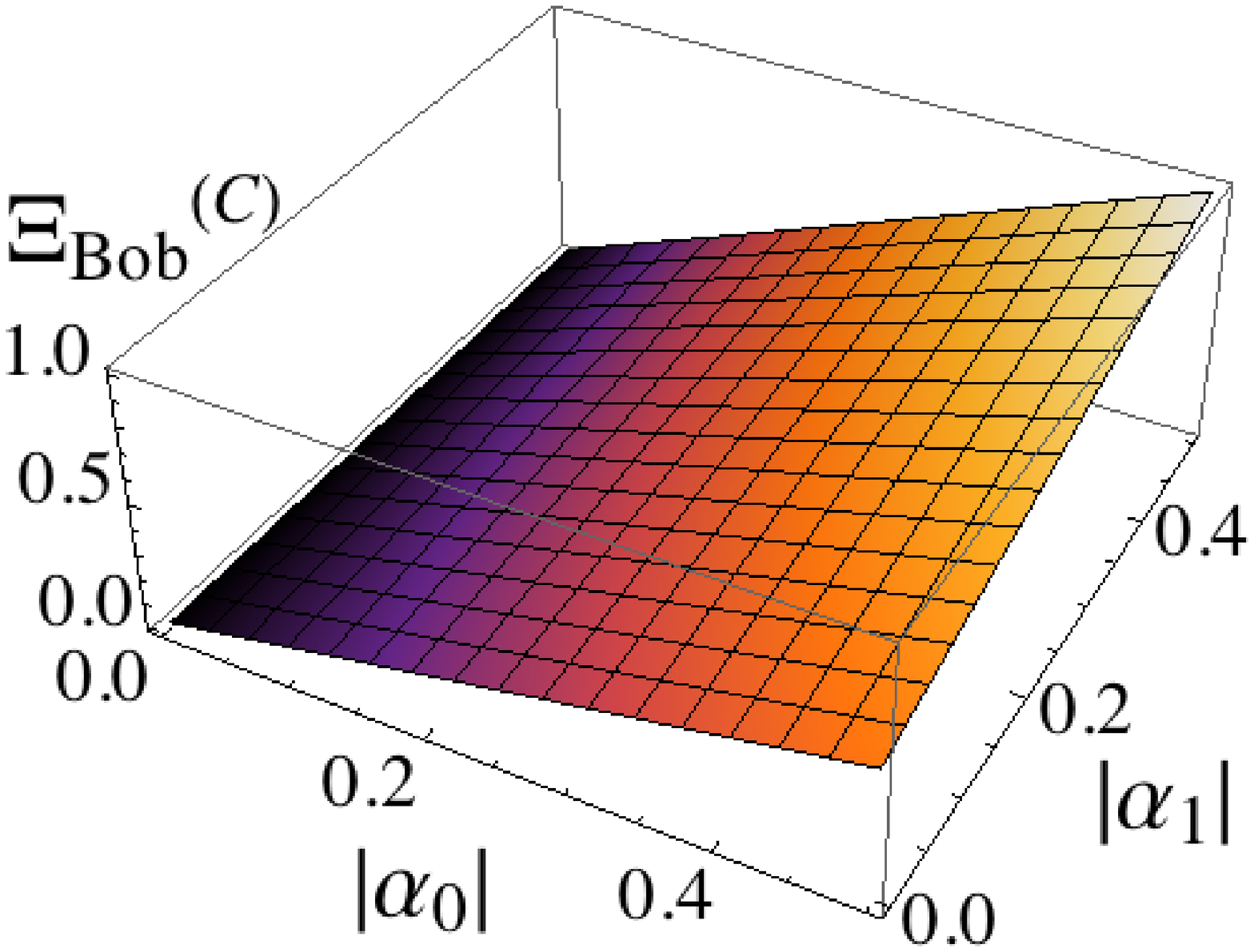}
\caption{(Color online) Bob's average score $\Xi_\text{Bob}^{(C)}$ (density-plot on the left, and 3D-plot on the right) with respect to $\abs{\alpha_0}$ and $\abs{\alpha_1}$. Here we assume that Bob's brain performs the reasoning with the well-quantified probabilities of his preferences.}
\label{fig:Payoff_c}
\end{figure}

$1$) {\em Average score achievable from the classical probabilistic reasoning.} -- First, we write out explicitly the conditional probabilities $\text{Pr}(u_\text{Bob}|x)$ ($x = 0,1$) in Eq.~(\ref{eq:payoff_av_tau}):
\begin{eqnarray}
\text{Pr}(u_\text{Bob}=0|x=0) &=& \frac{1}{2} + \alpha_0, \nonumber \\
\text{Pr}(u_\text{Bob}=1|x=0) &=& \frac{1}{2} - \alpha_0, \nonumber \\
\text{Pr}(u_\text{Bob}=0|x=1) &=& \frac{1}{2} + \alpha_0 \alpha_1, \nonumber \\
\text{Pr}(u_\text{Bob}=1|x=1) &=& \frac{1}{2} - \alpha_0 \alpha_1.
\label{eq:Prob_hC}
\end{eqnarray}
Here, from Eqs.~(\ref{eq:payoff_gen})-(\ref{eq:payoff_av_tau}), we can derive that, if there is no biased value of the factor $\boldsymbol\alpha$ (i.e., $\alpha_0=\alpha_1=0$), and hence Bob's preferences, then Bob's total average score $\Xi_\text{Bob}$ will be $0$. In such a case, Bob would become {\em indifferent} (i.e., `$\sim$') to the choice of his strategies. However, if Bob can have a finite non-zero value of $\boldsymbol\alpha$ for the given hints, the winning average can be improved. For example, when Alice's strategies $u_\text{Alice}(x)$ are [$\boldsymbol\tau.{\bf 1}$] and the probabilities of Bob's preferences are well-quantified with the directional condition ($\alpha_0 > 0$, $\alpha_1 > 0$) (as depicted in Fig.~\ref{fig:good_advice}), Bob's average score can be increased up to $\alpha_0 + 2\alpha_0\alpha_1 > 0$. By generalizing this advantage for other cases, it is found that Bob can have
\begin{eqnarray}
\Xi_\text{Bob}^{(C)} = \abs{\alpha_0} + 2\abs{\alpha_0}\abs{\alpha_1} > 0,
\label{eq:avS_C}
\end{eqnarray}
where the superscript `($C$)' means that the score is achievable from the classical probabilistic reasoning. In Fig.~\ref{fig:Payoff_c}, the graphs of $\Xi_\text{Bob}^{(C)}$ are given with respect to $\abs{\alpha_0}$ and $\abs{\alpha_1}$, assuming that the probabilities of Bob's preferences are well-quantified for the given hints. However, if Bob uses ill-quantified probabilities, Bob could have $\Xi_\text{Bob}^{(C)} < 0$, decreasing his winning average (see appendix~\ref{appendix:A} for details about this). This situation may arise when the hints are made with any malicious intention. 

\begin{figure}[t]
\centering
\includegraphics[width=0.26\textwidth]{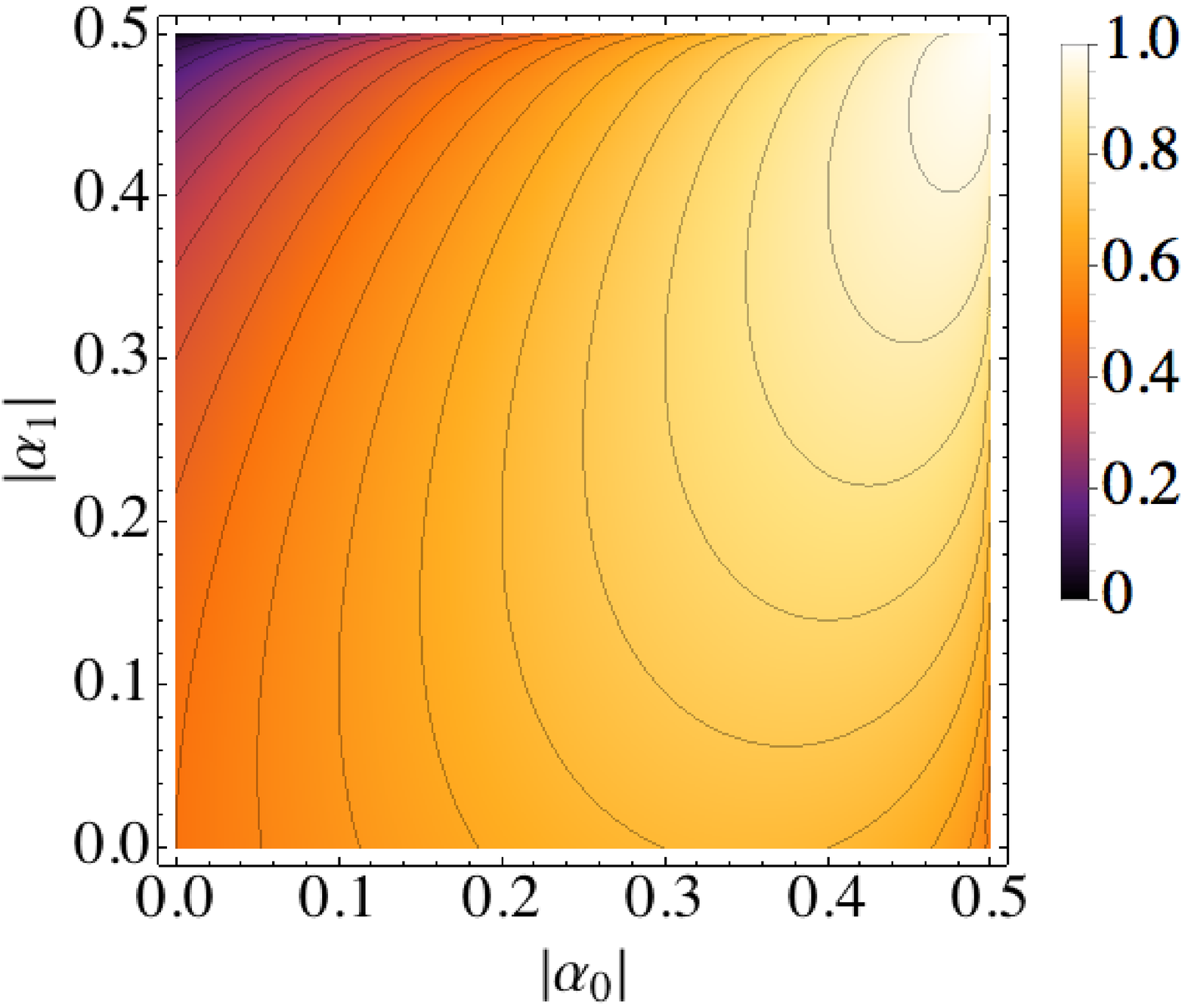}
\includegraphics[width=0.2\textwidth]{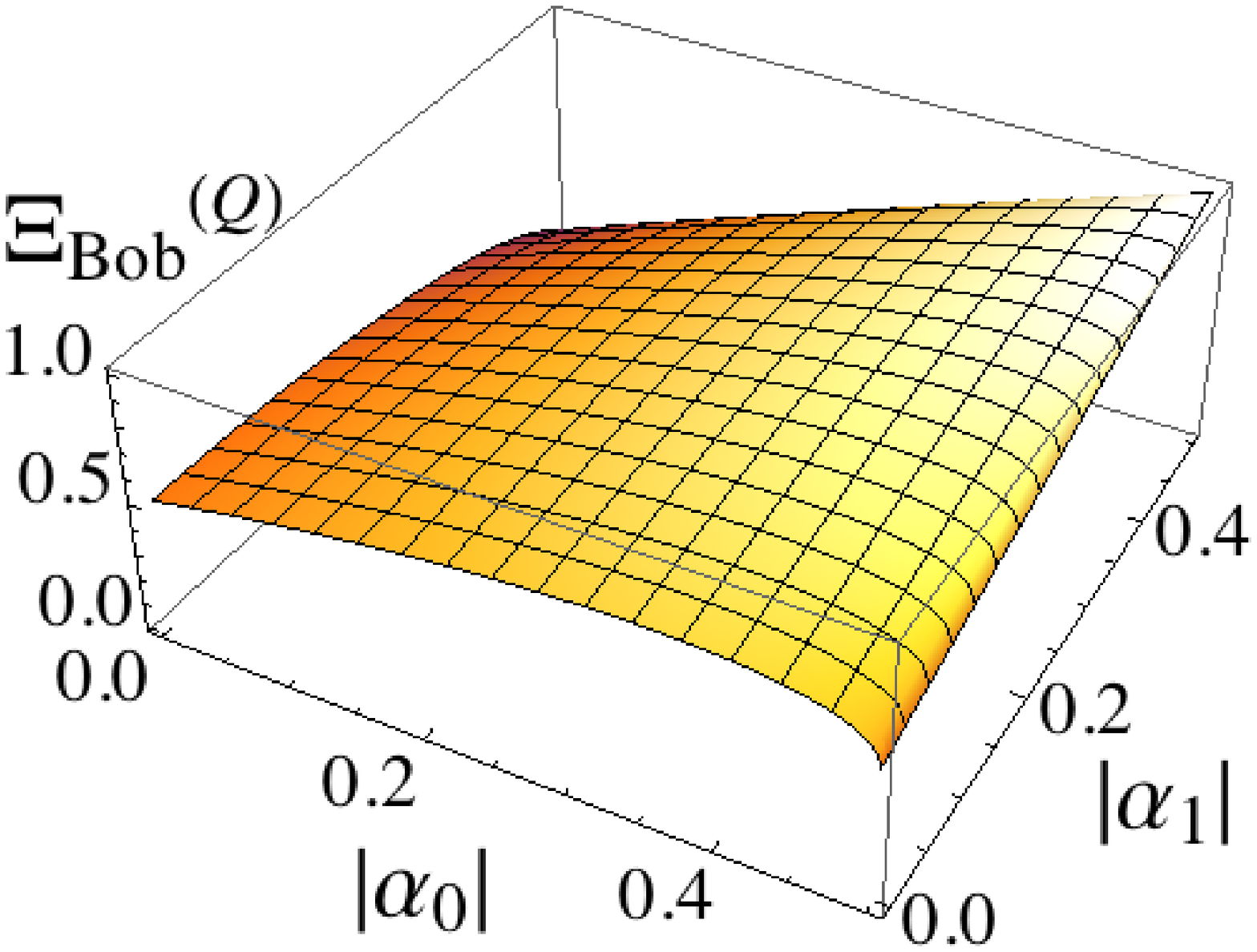}
\caption{(Color online) Bob's average score $\Xi_\text{Bob}^{(Q)}$ (density-plot on the left, and 3D-plot on the right). The probabilities as in Eq.~(\ref{eq:quantified_hint}) are also assumed to be well-quantified, and Bob can chose appropriate phase factors $\phi_j$ ($j=0,1$) following the rules in Eq.~(\ref{eq:Delta}).}
\label{fig:Payoff_q}
\end{figure}

$2$) {\em Average scores achievable from the quantum reasoning.} -- To analyze Bob's score achievable from the quantum reasoning, we also evaluate the conditional probabilities $\text{Pr}(u_\text{Bob}|x)$ ($x = 0,1$) as
\begin{eqnarray}
\text{Pr}(u_\text{Bob}=0|x=0) &=& \frac{1}{2} + \alpha_0, \nonumber \\
\text{Pr}(u_\text{Bob}=1|x=0) &=& \frac{1}{2} - \alpha_0, \nonumber \\
\text{Pr}(u_\text{Bob}=0|x=1) &=& \frac{1}{2} + \alpha_0 \alpha_1 + \Gamma \cos(\pi\Delta), \nonumber \\
\text{Pr}(u_\text{Bob}=1|x=1) &=& \frac{1}{2} - \alpha_0 \alpha_1 - \Gamma \cos(\pi\Delta),
\label{eq:Prob_hQ}
\end{eqnarray}
where $\Gamma$ is defined as
\begin{eqnarray}
\Gamma = 2\sqrt{\left(\frac{1}{4}-\abs{\alpha_0}^2\right)\left(\frac{1}{4}-\abs{\alpha_1}^2\right)}.
\label{eq:Gamma}
\end{eqnarray}
Here we readily see that the additional term ``$\Gamma\cos(\pi\Delta)$'' appears in the case of $x=1$. We note, again, that the factor $\Delta$ comes from the quantum phases $\phi_j$ ($j=0,1$) involved in the unitary gates $\hat{R}_j$ ($j=0,1$). Thus, by applying the rules of Eq.~(\ref{eq:Delta}), Bob can get (as long as $\alpha_0 \neq 0$ and $\alpha_1 \neq 0$)
\begin{eqnarray}
\Xi_\text{Bob}^{(Q)} = \Xi_\text{Bob}^{(C)} + \Gamma,
\label{eq:avS_Q}
\end{eqnarray}
where the superscript `($Q$)' denotes the score obtained by quantum reasoning. Here, by observing Eq.~(\ref{eq:avS_Q}), we can directly see that $\Xi_\text{Bob}^{(Q)}$ is always larger than or equal to $\Xi_\text{Bob}^{(C)}$, which means that Bob can increase his winning average more than in the classical probabilistic reasoning, {\em notably even in the case where the elements of the game are all classical}. The equality is satisfied when the given hints are perfect; namely, when $\abs{\alpha_0} = \abs{\alpha_1} = \frac{1}{2}$. This is quite natural, because if the given hints contain whole information of $u_\text{Alice}(x)$, then Bob can have  {\em strong preferences} (i.e., `$\succ$') toward his $100\%$ winning \footnote{One can also find that $\Xi_\text{Bob}^{(Q)} = \Xi_\text{Bob}^{(C)}$ when Bob's brain cannot determine the factor $\Delta$ with the condition of $\alpha_0 = 0$ or $\alpha_1 = 0$. But, this is trivial case. Note that Eq.~(\ref{eq:avS_Q}) is defined for $\alpha_0 \neq 0$ and $\alpha_1 \neq 0$ with the rule of Eq.~(\ref{eq:Delta}).}. In Fig.~\ref{fig:Payoff_q}, we give the graphs of $\Xi_\text{Bob}^{(Q)}$ for the well-quantified probabilities. However, we should point out that Bob's winning average can also be decreased due to any malicious hinting. In particular, $\Xi_\text{Bob}^{(Q)}$ could be much smaller than $\Xi_\text{Bob}^{(C)}$ in the worst case (see appendix~\ref{appendix:A}). 

\section{Numerical simulations}\label{sec:5}

\begin{figure}[t]
\centering
\includegraphics[width=0.23\textwidth]{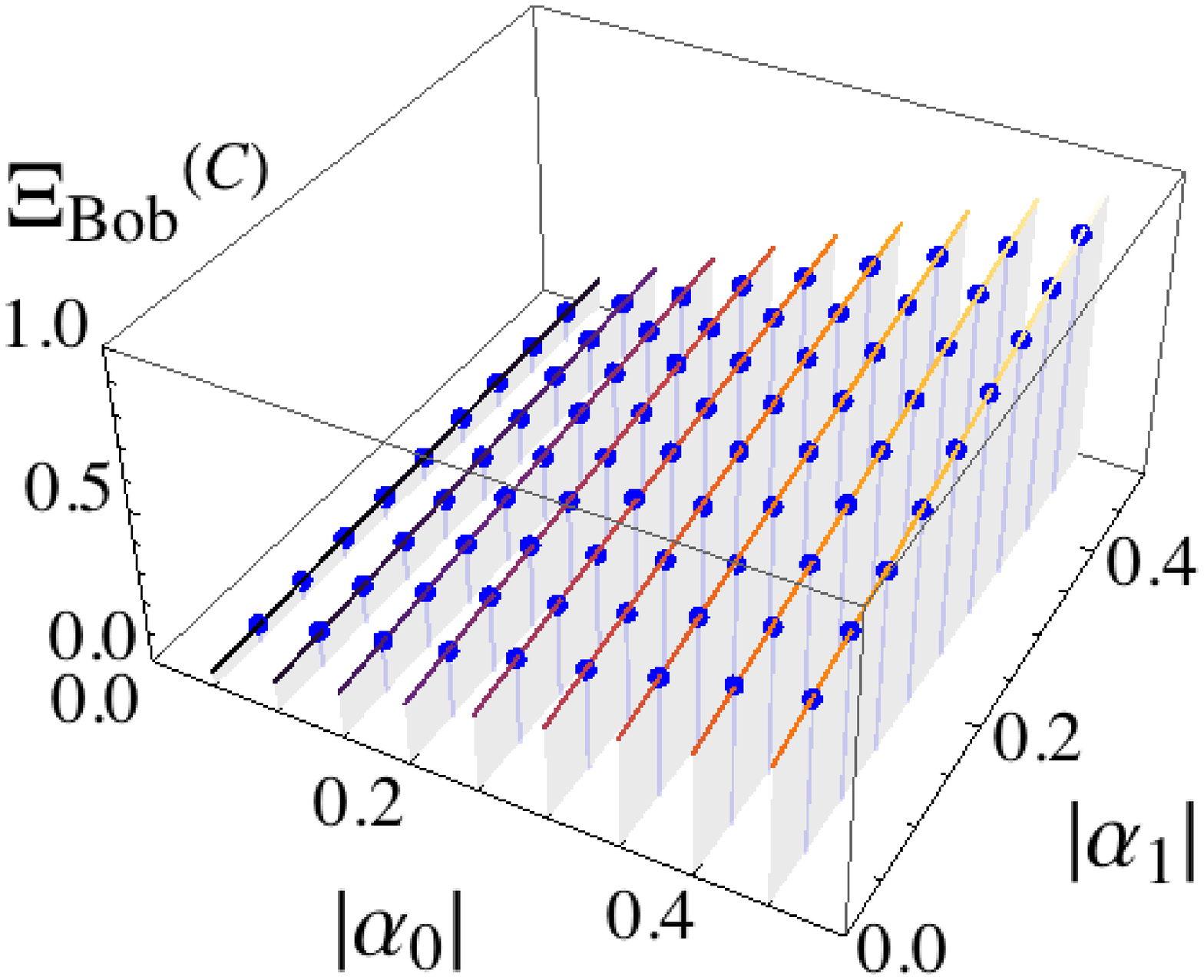}
\includegraphics[width=0.23\textwidth]{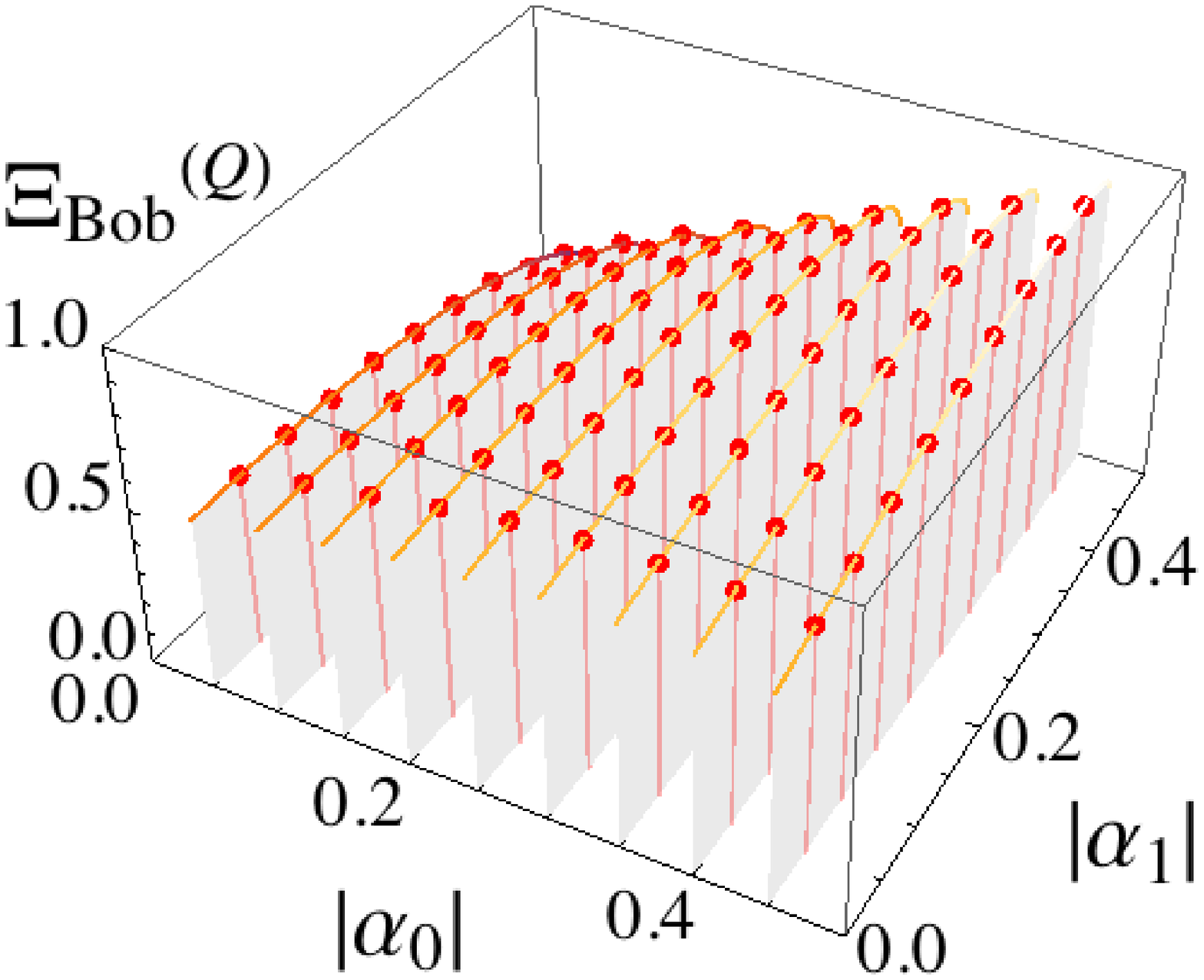}
\caption{(Color online) The simulation data of $\Xi_\text{Bob}^{(C)}$ (blue circle) and $\Xi_\text{Bob}^{(Q)}$ (red circle) are plotted for the two reasonings: (left) Classical probabilistic, and (right) quantum. The solid lines are the theoretical values drawn by Eq.~(\ref{eq:avS_C}) and Eq.~(\ref{eq:avS_Q}). The data are very well matched to the theoretical lines.}
\label{fig:Nsim_3D}
\end{figure}

We now demonstrate the results of our theoretical analysis through numerical simulations that are designed based on Fig.~\ref{fig:classical_a} and Fig.~\ref{fig:quantum_a}. Firstly, we assume that Bob enjoys a finite number of games $N$ following each of the two reasoning processes. Here, Alice chooses her secret-bits randomly in each game and Bob's brain always uses well-quantified probabilities (i.e., the certain values of $\abs{\alpha_0}$ and $\abs{\alpha_1}$ with appropriately assigned directional conditions). In Fig.~\ref{fig:Nsim_3D}, we plot the data of $\Xi_\text{Bob}^{(C)}$ and $\Xi_\text{Bob}^{(Q)}$ obtained from (left) classical probabilistic and (right) quantum reasoning. The data are plotted in the space of $\abs{\alpha_0}$ and $\abs{\alpha_1}$ (from $0.05$ up to $0.45$ at $0.05$ intervals). Each point is made by averaging the scores over $N=10^4$ games. Note that the graphs are duplicating Fig.~\ref{fig:Payoff_c} and Fig.~\ref{fig:Payoff_q}. Actually, the data are very well matched to the theoretical values (the solid lines) drawn by Eq.~(\ref{eq:avS_C}) and (\ref{eq:avS_Q}). 

\begin{figure}[t]
\centering
\includegraphics[width=0.46\textwidth]{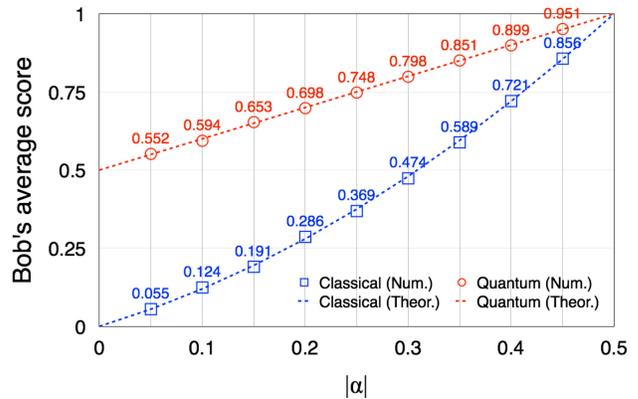}
\caption{(Color online) We present the data of $\Xi_\text{Bob}^{(C)}$ (blue squre) and $\Xi_\text{Bob}^{(Q)}$ (red circle), by assuming that $\abs{\alpha} = \abs{\alpha_0} = \abs{\alpha_1}$. The quantified probabilities in Eq.~(\ref{eq:quantified_hint}) are assumed to be good. Each data point is also averaged over $10^4$ trials of the game. Here we can see that the data are also very well matched to the theoretical (blue and red dashed) lines, drawn by Eq.~(\ref{eq:avS_C}) and Eq.~(\ref{eq:avS_Q}).}
\label{fig:N_sim}
\end{figure}

We then plot the data of $\Xi_\text{Bob}^{(C)}$ (blue square) and $\Xi_\text{Bob}^{(Q)}$ (red circle) with respect to $\abs{\alpha}$ in Fig.~\ref{fig:N_sim}. Here, we let $\abs{\alpha}=\abs{\alpha_0}=\abs{\alpha_1}$. Each data point is also averaged over $N=10^4$ games, and the quantified probabilities in Eq.~(\ref{eq:quantified_hint}) are assumed to be good. The dashed (blue and red) lines denote the theoretical values, drawn by Eq.~(\ref{eq:avS_C}) and Eq.~(\ref{eq:avS_Q}). In this case, it is directly seen that the increments of Bob's average scores from the quantum reasoning are higher than those from the classical probabilistic reasoning. Notably, degree of the increment is conspicuous when the amount of the Bob's preferences are very weak (as long as $\abs{\alpha} \neq 0$). Actually, Bob can increase his average scores more than $0.5$ when $\abs{\alpha}=0.05$ from the quantum reasoning, whereas the increments allowed from the classical probabilistic reasoning are vanishingly small. Note that $\Xi_\text{Bob}^{(C)}=\Xi_\text{Bob}^{(Q)}=0$ when $\abs{\alpha}=0$.

\section{Summary and discussion}

In summarizing, we have presented a classical two-player (Alice and Bob) game, called the Secret-Bit Guessing Game, where Bob attempts to guess what Alice's bits are. Using this game, we designed a legitimate process of Bob's reasoning using a simple Boolean function and defined one-player (Bob's) reasoning problem in the context of the theory of decision-making. We then considered two parallel ways of Bob's reasoning: One is classical probabilistic, and the other is quantum. We primarily investigated whether or not Bob can get the quantum advantage, particularly without changing the classical setting of the game. We replaced each reasoning which may occur in Bob's brain to a machinery process with the corresponding internal devices. On the basis of the analysis of payoff function, we explicitly showed that Bob can make better use of his weak preferences with quantum reasoning, faithfully dealing with quantum superposition. This quantum advantage was possible because the main logical operations present in Bob's brain provided another degree of freedom due to the quantum phase, and this enabled the {\em rational} player, Bob, to explore an additional way of using his weak preferences to maximize his chance of winning. The important scientific message of our study is: {\em It appears to be possible to get a quantum advantage even in the case where all strategies are classical}. We additionally investigated (in appendix~\ref{appendix:A}) that if the hints are made to deceive Bob, then Bob's winning average can decrease in general. However, in the worst case, such a disadvantage becomes much more acute. Thus, the quantum advantage in our game was counter-balanced with malicious hinting, and allowed us to remark on how to maximize the potential quantum advantages in such a system (see also Ref.~\cite{Anand14}).

As a response, one may consider that Bob's brain can probabilistically simulate the single-qubit process with the classical stuffs, and duplicate the measurement outcomes which accurately compare with those from the quantum reasoning \footnote{Actually, one may simulate the quantum reasoning process spending a finite additional computational resources (e.g., see Ref.~\cite{Chappell09} and Ref.~\cite{Balakrishnan13})}. However, this does not mean ``there is nothing the quantum'' \footnote{For example, see the comment in Ref.~\cite{SJvanEnk00}, and the reply in Ref.~\cite{MayerReplies00}.}. In fact, to argue that ``a single-qubit cannot be viewed as a genuine quantum system'' as ``it can classically be simulable'' is a long-standing problem, and for several years studies have shown that the single-qubit is incompatible with classical models in terms of temporal inequalities \cite{Ruskov06,Jordan06,DeZela07}, no-go theorems \cite{Cabello03,Grudka08}, operational quasi-probability \cite{Ryu13}, etc.

We believe that our work can provide some intuition on how can we get a quantum advantage using classical information or classical data. This question is of particular significance, since it may be related to some recent issues, e.g., in the field of quantum machine learning algorithm (see Ref.~\cite{Aaronson15} for more details). Our work is also expected to open up follow up studies across multiple disciplines, such as quantum cryptography and artificial intelligence.

\section*{Acknowledgments}

JB thanks professor Wonmin Son. We acknowledge the financial support of the Basic Science Research Program through the National Research Foundation of Korea (NRF) grant funded by the Ministry of Science, ICT \& Future Planning (No. 2014R1A2A1A10050117) and also the ICT R\&D program of MSIP/IITP (No. R0190-15-2030). MP and JR are supported by TEAM project of FNP. JR also acknowledges support by EU grant BRISQ2.

\appendix

\section{Further analyses: `well-quantified' and `ill-quantified' probabilities of Bob's preferences}\label{appendix:A} 

\begin{figure}[t]
\centering
\includegraphics[width=0.23\textwidth]{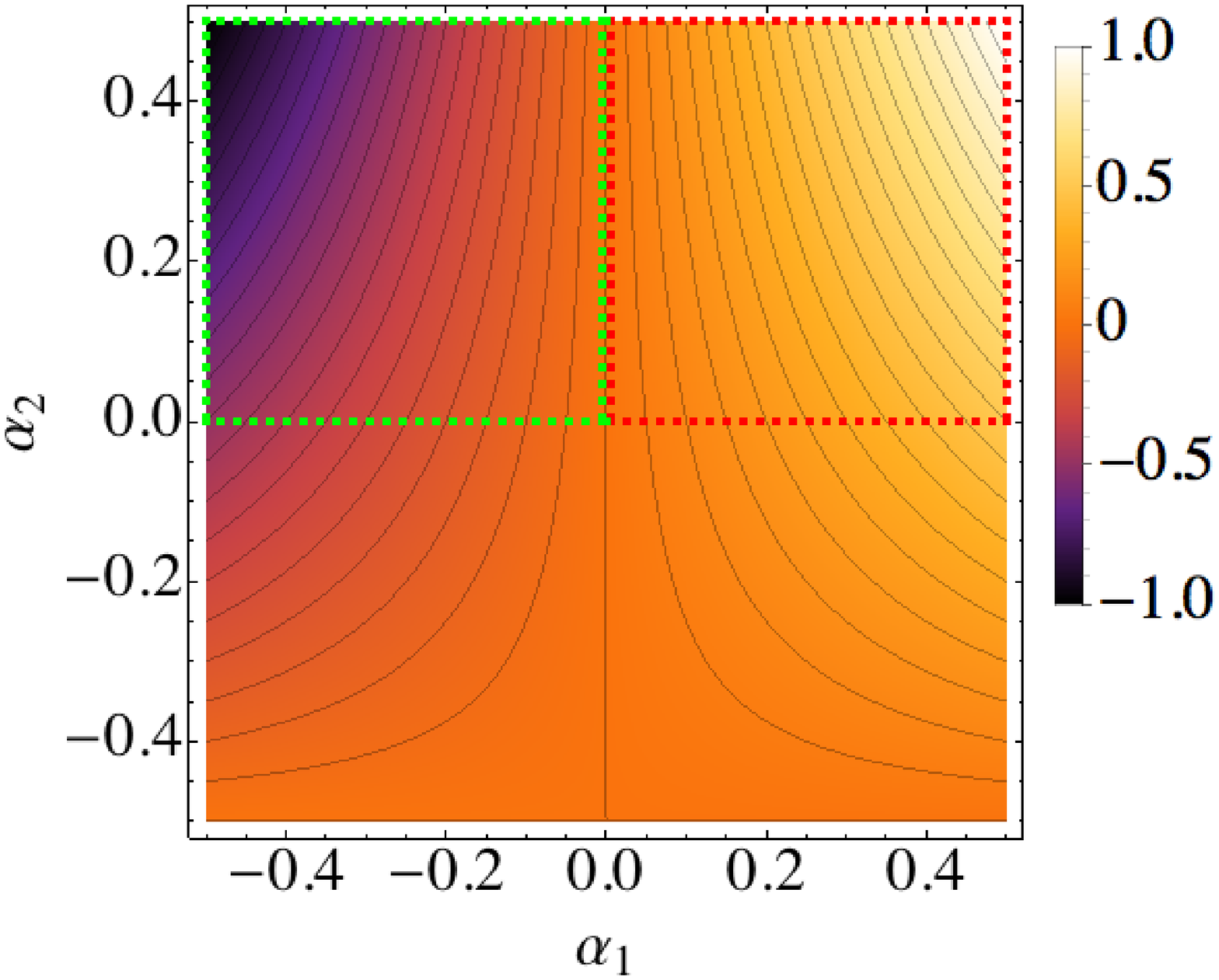}
\includegraphics[width=0.23\textwidth]{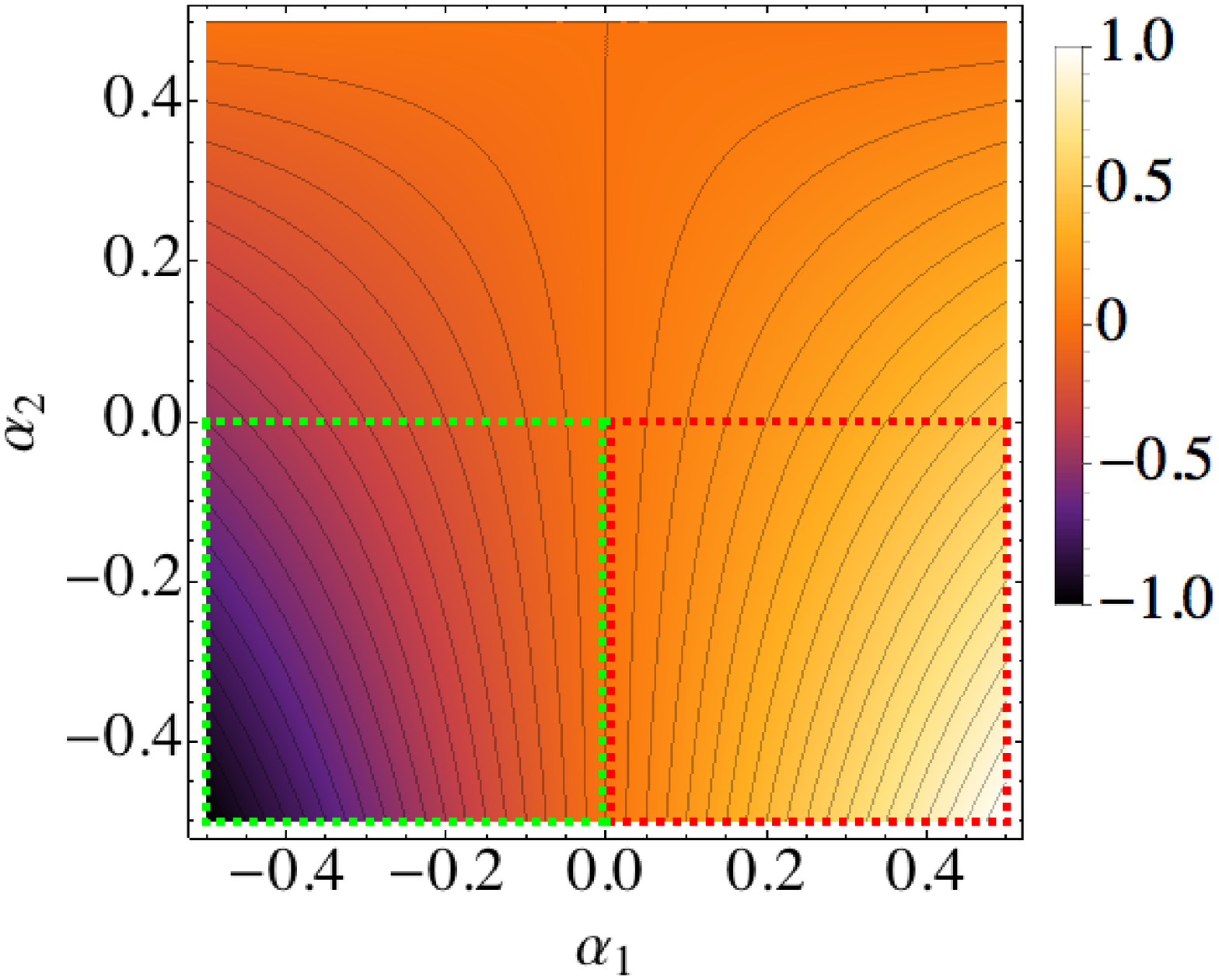}
\\
\includegraphics[width=0.23\textwidth]{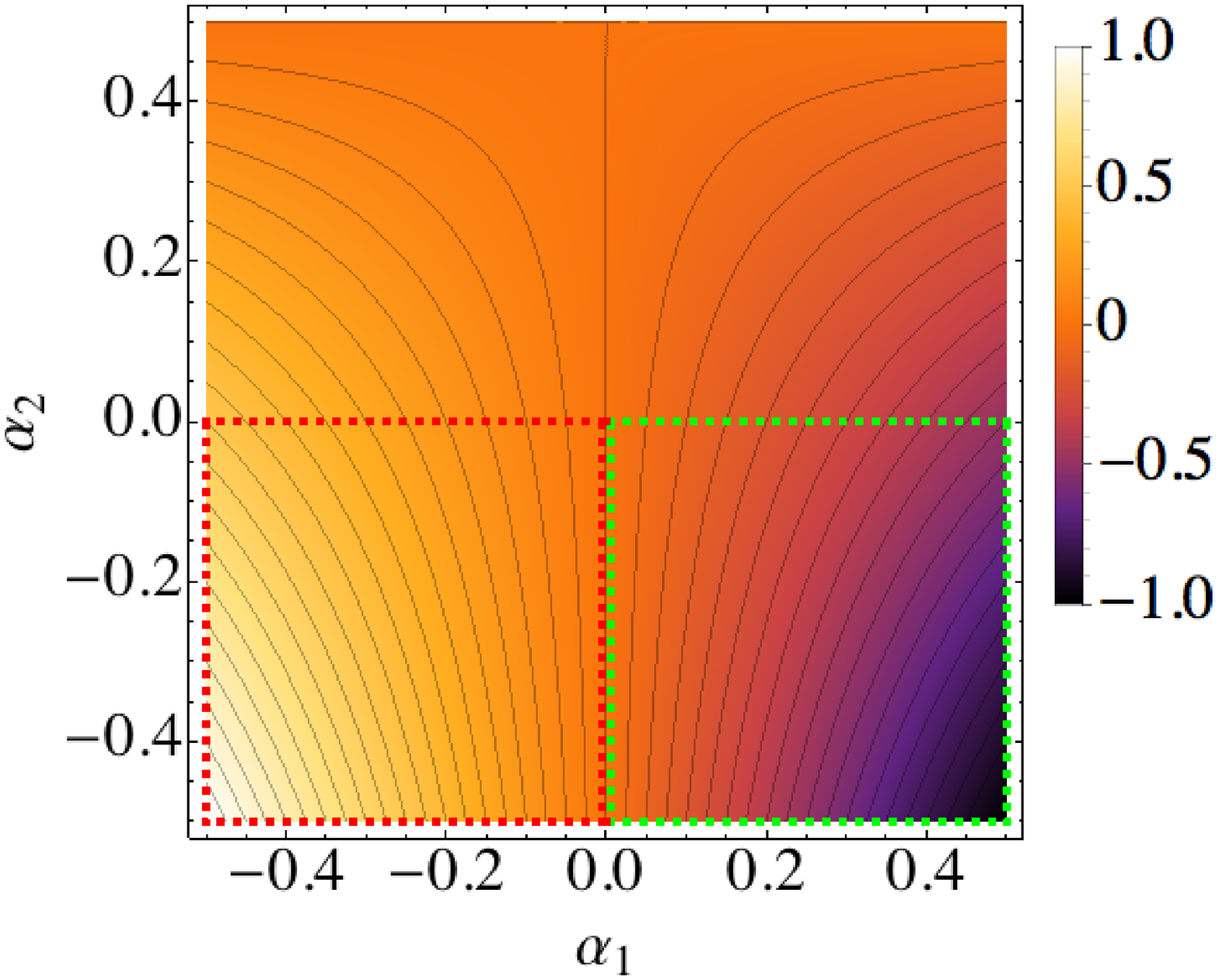}
\includegraphics[width=0.23\textwidth]{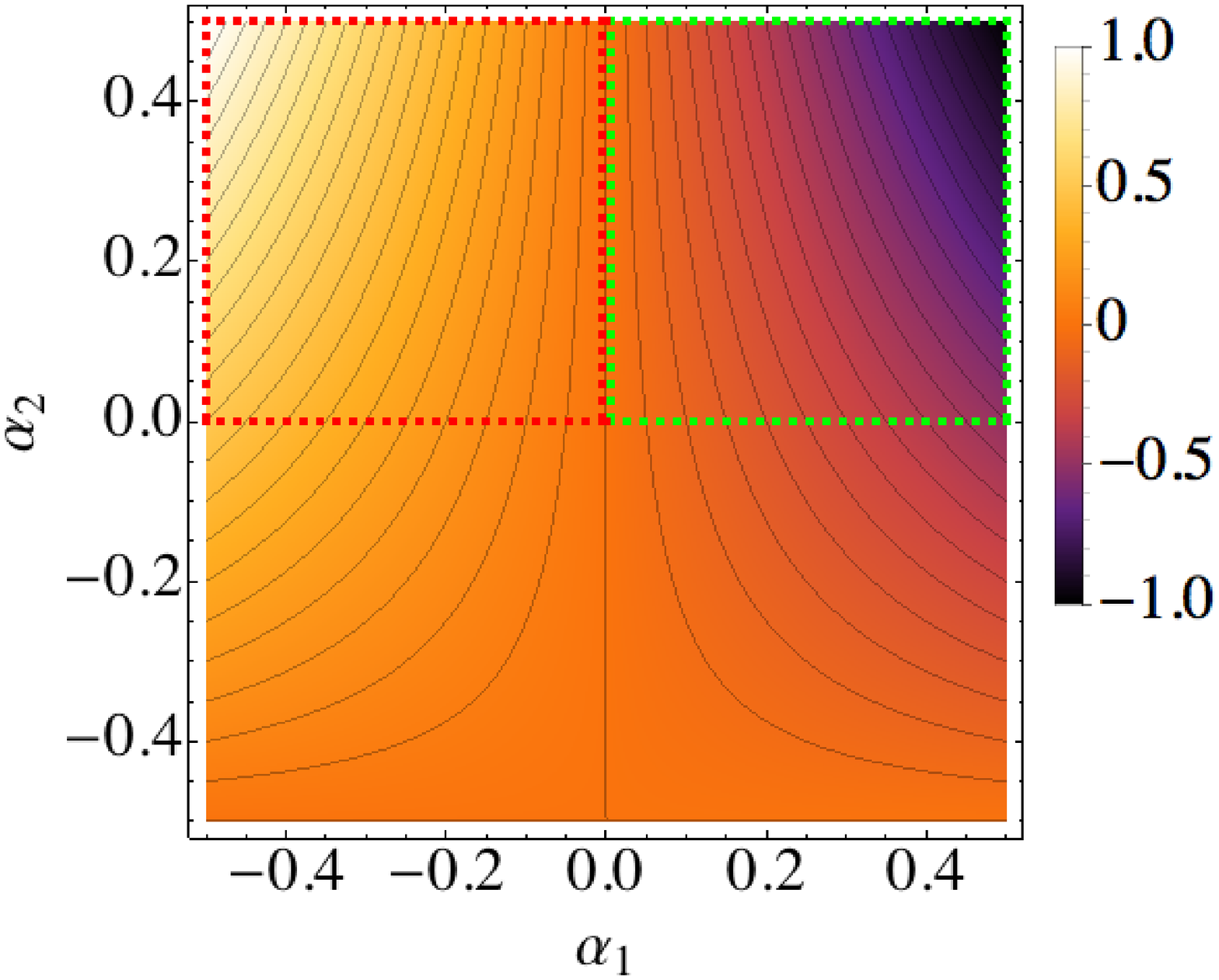}
\caption{(Color online) We depict Bob's score $\overline{\xi}_{\text{Bob},\tau}^{(C)}$ averaged for a specific set of $u_\text{Alie}(x)$ ($x=0,1$): (top-left) [$\boldsymbol\tau.{\bf 1}$] (top-right) [$\boldsymbol\tau.{\bf 2}$], (bottom-left) [$\boldsymbol\tau.{\bf 3}$], and (bottom-right) [$\boldsymbol\tau.{\bf 4}$]. We specify the regions of `good' (red dashed box) and `bad' (green dashed box) probabilities (see, also, Fig.~\ref{fig:good_advice}).}
\label{fig:app_score_Cl}
\end{figure}

Here we give further analyses for well-quantified and ill-quantified probabilities of Bob's preferences. Firstly, we recall the total average score $\Xi_\text{Bob}^{(C)}$ of Bob achievable from the classical probabilistic reasoning. From Eq.~(\ref{eq:payoff_av_tau}) and Eq.~(\ref{eq:Prob_hC}), we write $\overline{\xi}_{\text{Bob},\tau}^{(C)}$ ($\tau=1,2,3,4$) as below.
\begin{eqnarray}
\overline{\xi}_{\text{Bob},1}^{(C)} &=& \alpha_0 + 2\alpha_0 \alpha_1, \nonumber \\
\overline{\xi}_{\text{Bob},2}^{(C)} &=& \alpha_0 - 2\alpha_0 \alpha_1, \nonumber \\
\overline{\xi}_{\text{Bob},3}^{(C)} &=& -\alpha_0 + 2\alpha_0 \alpha_1, \nonumber \\
\overline{\xi}_{\text{Bob},4}^{(C)} &=& -\alpha_0 - 2\alpha_0 \alpha_1.
\label{eq:pc_x}
\end{eqnarray}
It is clear that if there is no bias among the preferences, i.e., $\alpha_0=\alpha_1=0$, then $\overline{\xi}_{\text{Bob},\tau}^{(C)}=0$ for all $\tau=1,2,3,4$. However, if Bob has $\mathbf{\alpha}=(\alpha_0, \alpha_1)^T \neq (0,0)^T$ for the given hints, Bob can improve his winning average with the appropriately assigned directional conditions of $\mathbf{\alpha}$ (see Fig.~\ref{fig:good_advice}) . More specifically, Bob can have
\begin{eqnarray}
\Xi_\text{Bob,best}^{(C)} = \abs{\alpha_1} + 2 \abs{\alpha_1} \abs{\alpha_2} > 0,
\label{eq:xiC_best}
\end{eqnarray}
as described in the main text [see Eq.~(\ref{eq:avS_C})]. However, if the hinting is malicious, Bob may fail. To see this clearly, we draw the graphs of $\overline{\xi}_{\text{Bob},\tau}^{(C)}$ for the cases $\tau=1,2,3,4$ (see Fig.~\ref{fig:app_score_Cl}). In each graph, we specify the regions of well-quantified (red dashed box) and ill-quantified (green dashed box) probabilities in the space of ($\alpha_0$, $\alpha_1$). Here we can imagine any malicious hinting that misleads Bob toward the green dashed regions. This is the worst scenario for Bob, in which he will have the score
\begin{eqnarray}
\Xi_{\text{Bob,worst}}^{(C)} = -\abs{\alpha_1} - 2 \abs{\alpha_1} \abs{\alpha_2}.
\label{eq:xiC_worst}
\end{eqnarray}
If Bob has to take into account all these situations, it is evident that Bob will have $\Xi_\text{Bob}=0$, because he cannot have any preferences for the given hints.

\begin{figure}[t]
\centering
\includegraphics[width=0.23\textwidth]{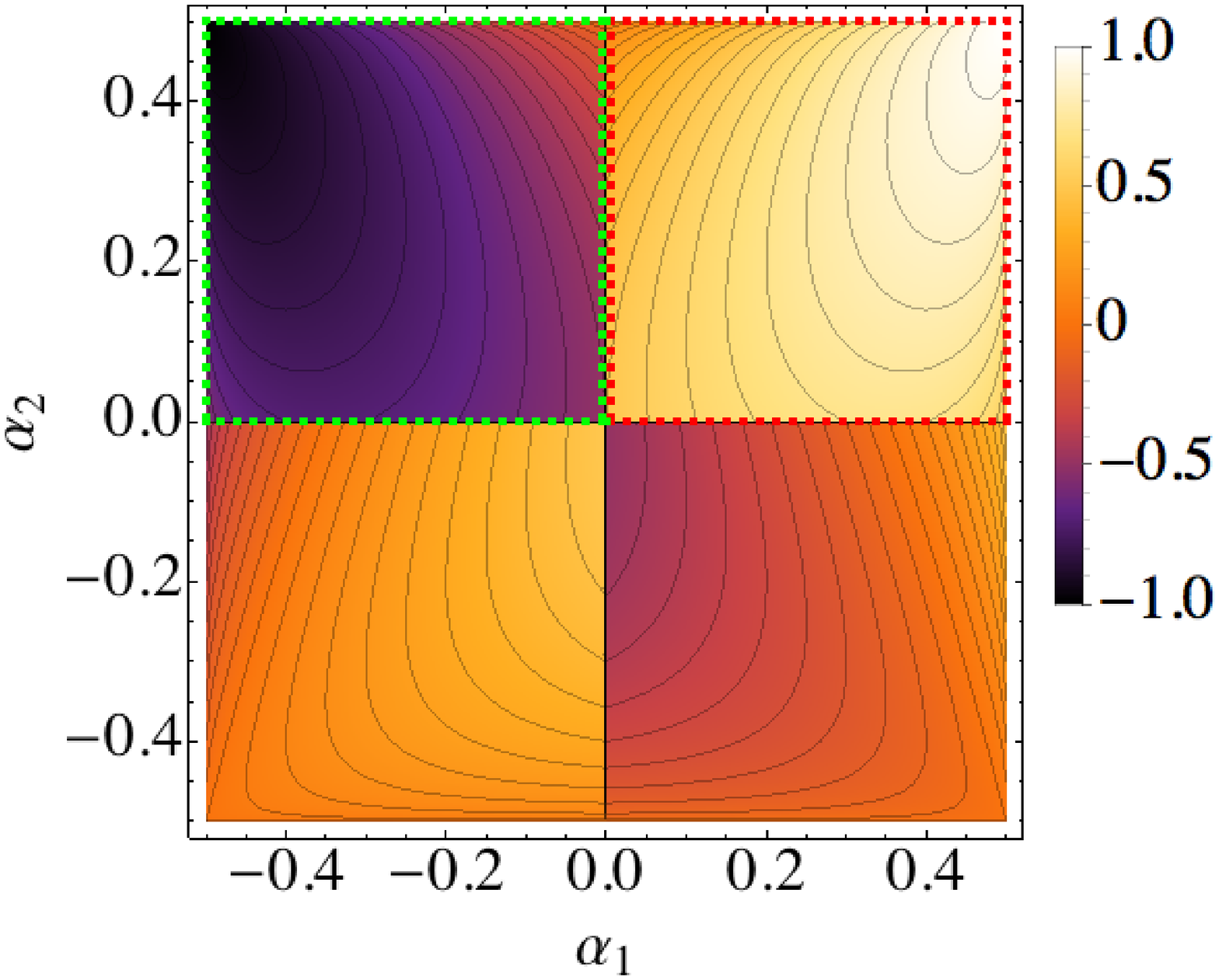}
\includegraphics[width=0.23\textwidth]{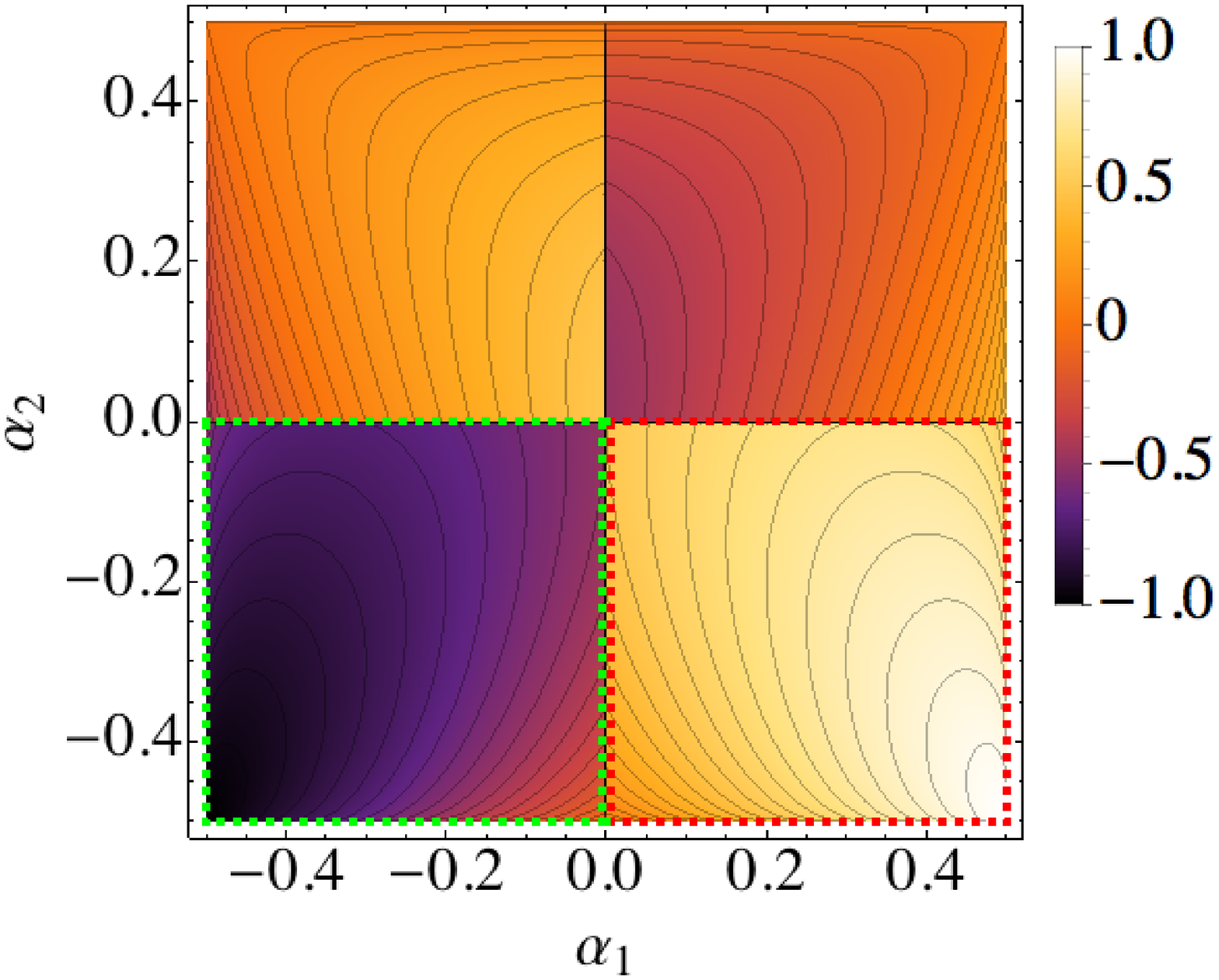}
\\
\includegraphics[width=0.23\textwidth]{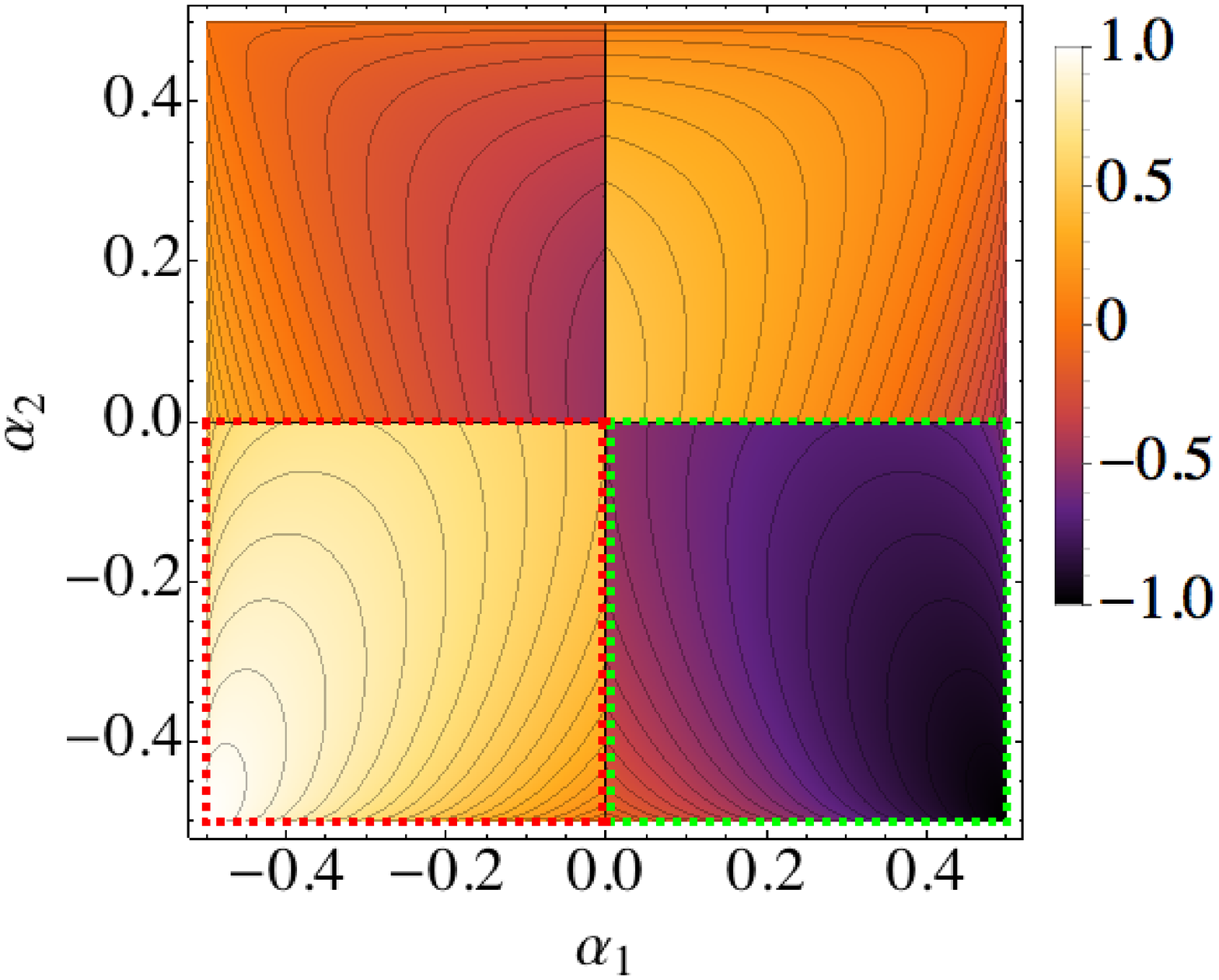}
\includegraphics[width=0.23\textwidth]{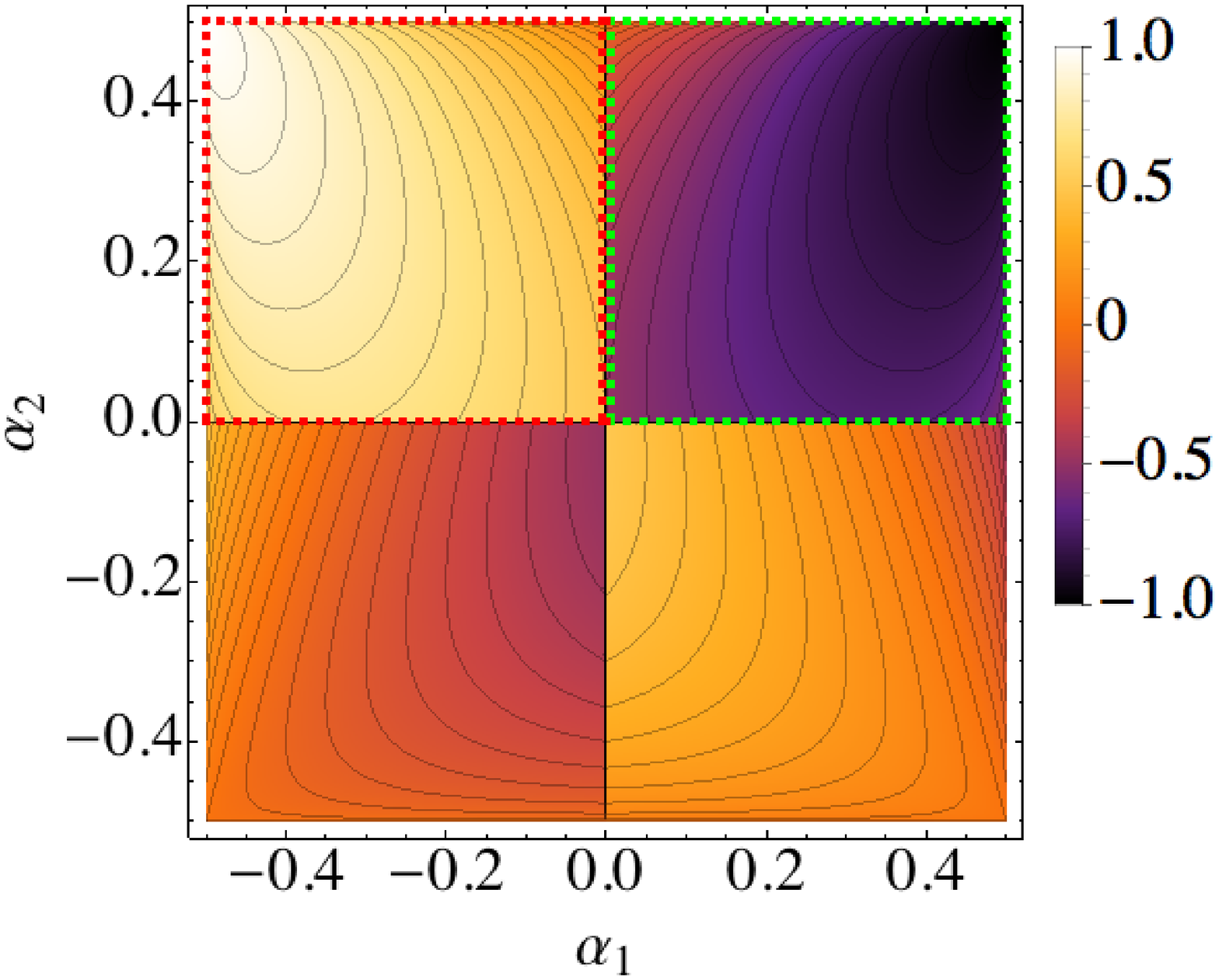}
\caption{(Color online) We give the graphs of $\overline{\xi}_{\text{Bob},\tau}^{(Q)}$ for (top-left) [$\boldsymbol\tau.{\bf 1}$], (top-right) [$\boldsymbol\tau.{\bf 2}$], (bottom-left) [$\boldsymbol\tau.{\bf 3}$], and (bottom-right) [$\boldsymbol\tau.{\bf 4}$]. We also specify the regions of `good' (red dashed box) and `bad' (green dashed box) probabilities.}
\label{fig:app_score_Qu}
\end{figure}

Then, turning our analysis to the quantum reasoning, let us write $\xi_{\text{Bob},\tau}^{(Q)}$, by using Eq.~(\ref{eq:payoff_av_tau}) and Eq.~(\ref{eq:Prob_hQ}), as
\begin{eqnarray}
\overline{\xi}_{\text{Bob},1}^{(Q)} &=& \overline{\xi}_{\text{Bob},1}^{(C)} + \Gamma\cos{(\pi\Delta)}, \nonumber \\
\overline{\xi}_{\text{Bob},2}^{(Q)} &=& \overline{\xi}_{\text{Bob},2}^{(C)} - \Gamma\cos{(\pi\Delta)}, \nonumber \\
\overline{\xi}_{\text{Bob},3}^{(Q)} &=& \overline{\xi}_{\text{Bob},3}^{(C)} + \Gamma\cos{(\pi\Delta)}, \nonumber \\
\overline{\xi}_{\text{Bob},4}^{(Q)} &=& \overline{\xi}_{\text{Bob},4}^{(C)} - \Gamma\cos{(\pi\Delta)},
\label{eq:pq_x}
\end{eqnarray}
where $\Gamma = 2\sqrt{\left(\frac{1}{4}-\abs{\alpha_1}^2\right)\left(\frac{1}{4}-\abs{\alpha_2}^2\right)}$ [see Eq.~(\ref{eq:Gamma})]. Here, it is also true that Bob cannot improve his winning average when $\alpha_0 = \alpha_1 =0$. In such a case, Bob has $\overline{\xi}_{\text{Bob},\tau}^{(C)}=0$ with $\cos{(\pi\Delta)}=0$ for all $\tau=1,2,3,4$. However, if he can use well-quantified probabilities in his quantum reasoning, the average score can be higher than Eq.~(\ref{eq:xiC_best}). Specifically, Bob can have
\begin{eqnarray}
\Xi_\text{Bob,best}^{(Q)} = \Xi_\text{Bob,best}^{(C)} + \Gamma, 
\label{eq:xiQ_best}
\end{eqnarray}
as described in the main text [see Eq.~(\ref{eq:avS_Q})]. However, there can also be malicious hinting, in which case Bob may fail, similarly to the classical case. Let us see the graphs of $\overline{\xi}_{\text{Bob},\tau}^{(Q)}$ in Fig.~\ref{fig:app_score_Qu}, where the regions of well-quantified (red dashed box) and ill-quantified (green dashed box) probabilities are also specified. From the same analysis as in the case of the classical probabilistic reasoning, we can see that Bob's total average scores can be decreased. Notably, in the worst case, such disadvantages can be maximized as
\begin{eqnarray}
\Xi_\text{Bob,worst}^{(Q)} = \Xi_\text{Bob,worst}^{(C)} - \Gamma.
\label{eq:xiQ_worst}
\end{eqnarray}
This implies that the quantum reasoning can make the situation worse. In the case where Bob cannot evaluate whether the given hints are good or not, it is not possible for Bob to improve his winnings, as described above.


\end{document}